\documentclass[,final]{aipproc}
\usepackage{rimjournals}
\newcommand{\Msun}{\ifmmode\mbox{M}_{\odot}\else$\mbox{M}_{\odot}$\fi}
\newcommand{\Rsun}{\ifmmode\mbox{R}_{\odot}\else$\mbox{R}_{\odot}$\fi}
\newcommand{\erg}{\ifmmode\textrm{erg}\else$\textrm{erg}$\fi}
\def\lsim{\lower 2pt \hbox{$\, \buildrel {\scriptstyle <}\over{\scriptstyle \sim}\,$}}
\def\gsim{\lower 2pt \hbox{$\, \buildrel {\scriptstyle >}\over{\scriptstyle \sim}\,$}}
\def\lapp{\ifmmode\stackrel{<}{_{\sim}}\else$\stackrel{<}{_{\sim}}$\fi}
\def\gapp{\ifmmode\stackrel{>}{_{\sim}}\else$\stackrel{>}{_{\sim}}$\fi}
\layoutstyle{8x11double}
\begin{document}
\title{An RXTE Archival Search for Coherent X-ray Pulsations in LMXB 4U~1820$-$30}
\author{Rim Dib}{
  address={McGill University}
}
\author{Scott Ransom}{
  address={McGill University}
}
\author{Paul Ray}{
  address={Naval Research Laboratory}
}
 \author{Victoria Kaspi}{
  address={McGill University}
}
\begin{abstract}
As part of a large-scale search for coherent pulsations from LMXBs in the
{\em RXTE} archive, we have completed a detailed series of searches for
coherent pulsations of 4U~1820$-$30 --- an ultracompact LMXB with a binary
period of 11.4 minutes located in the globular cluster NGC~6624.  The small
binary period leads to a very high acceleration, so we used phase modulation
searches as well as acceleration searches to give significant sensitivity to
millisecond pulsations. We searched a total of 34 archival {\em RXTE}
observations, 32 of which had an on-source integration time longer than
10\,ks, and some of which were made consecutively which allowed us to
combine them.  While we found no pulsations, we have been able to place the
first stringent (95\% confidence) pulsed fraction limits of \lapp0.8\%
for all realistic spin frequencies (i.e. \lapp1\,kHz) and likely
companion masses (0.02\,\Msun $\leq M_c \leq$ 0.3\,\Msun).  By contrast all
five LMXBs known to emit coherent pulsations have intrinsic pulsed fractions
in the range 3\% to 7\% when pulsations are observed.
\end{abstract}
\maketitle
\section{Introduction}
One of the great scientific expectations when the Rossi X-ray Timing 
Explorer ({\em RXTE}) was launched in 1995 was the discovery of coherent
pulsations from low-mass X-ray binaries (LMXBs).
At present, only five accreting millisecond pulsars are known.
All five are faint transient sources where the pulsations at the spin period
of the pulsar were discovered during an outburst (see the contributions of D.
Chakrabarty and C. Markwardt to the proceedings of this conference). 
\par
There are many things that we can learn from searching for more examples of   
direct pulsations from LMXBs. For example, 
coherent pulsations give the precise rotation rate of the neutron star and
test the connection between recycled MSPs and LMXBs. They help
constrain the models of kHz QPOs and burst oscillations. 
Timing of the coherent pulses would also allow measurements of the accretion
torques, the orbital parameters, and the companion masses.  In addition,
along with X-ray and optical spectra, pulse timing can give information
about the compactness and the equation of state of the neutron star. 
Finally, setting stringent upper limits to coherent pulsations in several
sources in a variety of spectral states will impose constraints on the
possible mechanism for supressing coherent pulsations in these sources. For
all these reasons, we have started a large-scale search for coherent
pulsations from LMXBs.
\section{Characteristics of 4U~1820$-$30} 
The source 4U~1820$-$30 is an atoll LMXB in globular cluster NGC 6624. It
has an orbital binary period of 685~s (11.4~min) \cite{spw87}, the shortest
known binary orbital period in an LMXB.
4U~1820$-$30 undergoes a regular $\sim$176~day accretion cycle \cite{pt84}
switching between high and low luminosity states.
In the low state, regular Type~I bursts are seen $\pm$23
days around the minimum luminosity \citep{cg01}. 
In the low state, 4U~1820$-$30 has also shown an extremely energetic
superburst, likely due to deep ignition of a carbon layer \cite{sb02}. 
Several low frequency QPOs \cite{wvr99} as well as two peaks of kHz~QPOs
\cite{szw97} have been observed from this source. 
Faint UV and optical counterparts of 4U~1820$-$30 have also been observed
\cite{ksa+93,amd+97}. 
If the secondary star in the system is a white dwarf, its mass is estimated
to be 0.058 to 0.078~\Msun~\cite{rmj+87}. If the secondary is a main
sequence star, the upper limit on its mass is 0.3~\Msun~\cite{ak93}.  
\par
Of all known LMXBs, we searched 4U~1820$-$30 first because a lot of high
time resolution long data sets of this source are available in the archives,
because the presence of kHz QPOs in this source may indicate favorable
conditions for detecting pulsations, and because of its small orbital
period: the phase modulation search technique that we are using is most
sensitive (and provides the biggest increase in sensitivity over previous
methods) when the observations are longer than two complete binary orbits
\cite{rce03x}. 
\section{The {\em{RXTE}} Observations}
We searched 34 archival {\em{RXTE}} observations collected between 1996 and
2002.  The observations were available in event modes, and had a time
resolution of 125~$\mu$s or better. 32 of the observations had a total
time on source bigger than 10~ks.  The longest of these was 25~ks long with a
total time on source of 16~ks.  Some of the observations were segments of
longer observations which allowed us to concatenate them and analyse them
together.  The longest of the concatenated observations was 77.5~ks long
with a total time on source of 46.5~ks.
\section{The data Analysis}
Each observation was downloaded, filtered, barycentered, and then split into
four energy bands: Soft (2-5 keV), Medium (5-10 keV), Hard (10-20 keV), and
Wide (2-20 keV), in order to attempt to maximize the signal-to-noise ratio
of the unknown pulsations.  This processing used custom Python scripts that
call the standard FTOOLs.  The processed events were then binned into high
time resolution (either 0.122 ms or 0.244 ms) time series. Fast Fourier
Transforms of the four time series were then computed.  Acceleration
searches of FFTs of various durations were performed. `Phase-Modulation'
searches were conducted on each of the full duration FFTs. Finally,
candidates above our threshold were examined using brute-force folding
techniques to determine if they were true pulsations.
\section{The Search Techniques}
After we prepared and binned the data sets, we searched them for pulsations
in the Fourier domain using two types of searches which we summarize below.
\subsection{Acceleration Searches} 
This is the method traditionally used to compensate for the effects of
orbital motion \cite{mk84,agk+90,jk91,whn+91x,vvw+94x}: A time series is
stretched or compressed appropriately to account for a trial constant
frequency derivative (i.e. constant acceleration), Fourier transformed, and
then searched for pulsations. We used a Fourier-domain variant of this
technique \cite{rem02x}. Unfortunately acceleration searches can detect only
the strongest pulsations from systems like 4U~1820$-$30 where the orbital
period is short. So while we did not expect acceleration searches to find
pulsations, we still used them because they are included in the standard
search pipeline that we are going to use to search the other sources.
\subsection{Phase Modulation Searches} 
In order to analyze longer observations we used a new `phase-modulation' or
`sideband' search technique \cite{rce03x}.  This technique relies on the
fact that if the observation time is longer than the orbital period, the
orbit phase-modulates the pulsar's spin frequency: phase modulation results
in a family of evenly spaced sidebands in the frequency domain, around the
intrinsic pulsar signal. The constant spacing between the sidebands is
related to the binary orbital period.
\par
A phase modulation search is conducted by taking short FFTs of the power
spectrum of a full observation. The short FFTs cover overlapping portions of
various lengths (to account for the unknown semi-major axis of the LMXB)
over all the Fourier frequency range of the original power spectrum (to
account for the unknown pulsation period). They are then searched for peaks
indicating regularly spaced sideband (to detect the orbital period of the
LMXB). A detection provides initial estimates of the pulsar period, the
orbital period, and the projected radius of the orbit, which are refined by
generating a series of complex-valued template responses to correlate with
the original Fourier amplitudes (i.e. matched filtering in the Fourier
domain) \cite{rce03x}.
\begin{figure}[ht]
\caption{Monte-Carlo derived sensitivity calculations for the pulsation
searches of archival RXTE observations of 4U~1820$-$30. The plotted
sensitivities are 95\% confidence limits using phase modulation search
technique for a typical RXTE observation assuming a pulsar spin period of
3ms. We would easily have detected any coherent pulsations with a pulsed
fraction~>~0.8\% for all realistic companion masses and spin periods.}
\includegraphics[height=.5\textheight]{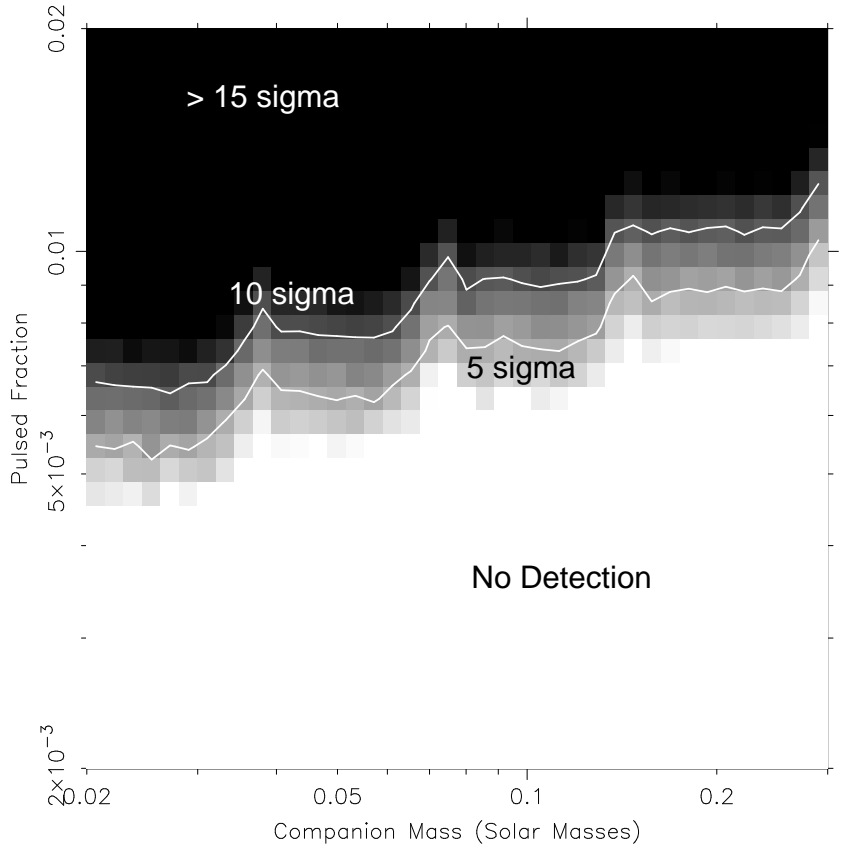}
\end{figure}
\section{Results and Upper Limits} 
Our searches did not detect pulsations. 
\par
To set an upper limit on the pulsed fraction of detectable pulsations from
4U~1820$-$30 we used the following procedure: we ran our searches on several
data sets containing simulated pulsations of various pulsed fractions for a
range of companion masses (see Figure 1) and for several spin frequencies.
For the purpose of the simulations, we assumed an orbital period of 685
seconds, the typical length of a long RXTE observation (20.5~ks), an inclination
angle of 60 degrees, and a circular orbit. Of the searches that we ran,
phase modulation searches run on the full duration data sets were the most
sensitive. Figure 1 shows the results of the simulations for this type of
search run on data sets containing fake 3~ms pulsations. Every point in the
figure corresponds to 100 searched data sets. The shading of every point
corresponds to a sigma detection level. The line of sigma = 5 indicates the
value of our upper limit on the pulsed fraction for a signal period of 3~ms.
This value varied between 0.55\% and 1.0\% (background subtracted) over the
range of companion masses that we used. The value decreased slightly when
the period of the signal was larger by a few milliseconds.  This means that
if there were a 3~ms coherent pulsation coming from 4U~1820$-$30 with a
pulsed fraction equal or higher than the upper limit stated above, our
searches would have easily (with a 95\% confidence) detected it. 
\par
Our upper limit on the pulsed fraction of possible signals from 4U~1820$-$30
was about 0.8\% (background subtracted) for spin periods under 10~ms.  The
pulsed fractions of the signals detected from the 5 LMXBs with known signal
period ranged between 3\% and 7\%, for signal frequencies between 2~ms and
6~ms \cite{source2, source3, source4, source3sum}. This means that our 
searches would have detected the pulsations from 4U~1820$-$30 if they were as
strong as the pulsations detected from the other 5 sources and in the same
period range.
\section{Potential reasons for the Lack of Observed Pulsations}
Several theories exist to explain why pulsations with a higher pulsed
fraction than our upper limit are not seen. 
To begin with, an unfavorable rotational geometry, an unfavorable
viewing geometry, or both can make pulsations undetectable. The pulsed
fraction may be reduced due to scattering in the surrounding medium
\cite{THEwhypaper}. Gravitational lens effects on the emission from the hot
polar caps can also greatly reduce the pulsed fraction
\cite{lensing1,lensing2}. Finally, screening of the stellar magnetic field
by the accreting matter can lead to a less preferential heating of the
surface (i.e. no polar caps) and therefore to the absence of observable
pulsations \cite{magpaper}.
\begin{theacknowledgments}
This work was supported by the Natural Sciences and Engineering Research   
Council (NSERC) Julie Payette research scholarship. It would also have been   
impossible to do this work without the computing power of the Beowulf 
computer cluster operated by the McGill University Pulsar Group and funded 
by the Canada Foundation for Innovation.
\end{theacknowledgments}

\end{document}